%% file: paper.tex
\documentclass[sigconf]{acmart}

\usepackage{glossaries}
\usepackage{balance}
\usepackage{listings}
\usepackage{xcolor}
\usepackage{csquotes}
\usepackage{enumitem}
\usepackage[framemethod=TikZ]{mdframed}
\usepackage{subcaption}
\usepackage{tabularx}
\usepackage{booktabs}
\usepackage{cleveref}
\usepackage{pifont}
\usepackage[htt]{hyphenat}

\newmdenv[backgroundcolor=gray!20, roundcorner=5pt, linewidth=1pt, skipabove=1em]{rqanswer}

\newlist{questions}{enumerate}{2}
\setlist[questions,1]{label=\textbf{RQ\arabic*},ref=RQ\arabic*}
\setlist[questions,2]{label= (\alph*),ref=\thequestionsi(\alph*)}

\colorlet{punct}{red!60!black}
\definecolor{delim}{RGB}{20,105,176}

\lstdefinelanguage{json}{
  basicstyle=\footnotesize\normalfont\ttfamily,
  numbers=left,
  numberstyle=\small,
  stepnumber=1,
  numbersep=8pt,
  showstringspaces=false,
  breaklines=true,
  frame=lines,
  literate=
    {:}{{{\color{punct}{:}}}}{1}
    {,}{{{\color{punct}{,}}}}{1}
    {\{}{{{\color{delim}{\{}}}}{1}
    {\}}{{{\color{delim}{\}}}}}{1}
    {[}{{{\color{delim}{[}}}}{1}
    {]}{{{\color{delim}{]}}}}{1},
}

\glsdisablehyper
\input{glossary.tex}

\AtBeginDocument{%
  }

\setcopyright{acmcopyright}
\copyrightyear{2018}
\acmYear{2018}
\acmDOI{XXXXXXX.XXXXXXX}

\acmConference[Conference acronym 'XX]{Make sure to enter the correct
  conference title from your rights confirmation emai}{June 03--05,
  2018}{Woodstock, NY}
\acmPrice{15.00}
\acmISBN{978-1-4503-XXXX-X/18/06}

\begin{document}

%%
%% The "title" command has an optional parameter,
%% allowing the author to define a "short title" to be used in page headers.
\title[You Can Run But You Can't Hide]{You Can Run But You Can’t Hide: Runtime Protection Against Malicious Package Updates For Node.js}

%%
%% The "author" command and its associated commands are used to define
%% the authors and their affiliations.
%% Of note is the shared affiliation of the first two authors, and the
%% "authornote" and "authornotemark" commands
%% used to denote shared contribution to the research.
\author{Marc Ohm}
\email{ohm@cs.uni-bonn.de}
\affiliation{%
  \institution{Fraunhofer FKIE \& University of Bonn}
  \streetaddress{Friedrich-Hirzebruch-Allee 8}
  \city{Bonn}
  \state{NRW}
  \country{Germany}
  \postcode{53115}
}

\author{Timo Pohl}
\email{pohl@cs.uni-bonn.de}
\affiliation{%
  \institution{University of Bonn}
  \streetaddress{Friedrich-Hirzebruch-Allee 8}
  \city{Bonn}
  \state{NRW}
  \country{Germany}
  \postcode{53115}
}

\author{Felix Boes}
\email{boes@cs.uni-bonn.de}
\affiliation{%
  \institution{University of Bonn}
  \streetaddress{Friedrich-Hirzebruch-Allee 8}
  \city{Bonn}
  \state{NRW}
  \country{Germany}
  \postcode{53115}
}

%%
%% By default, the full list of authors will be used in the page
%% headers. Often, this list is too long, and will overlap
%% other information printed in the page headers. This command allows
%% the author to define a more concise list
%% of authors' names for this purpose.
\renewcommand{\shortauthors}{Ohm et al.}

%%
%% The abstract is a short summary of the work to be presented in the
%% article.
\begin{abstract}
  Maliciously prepared software packages are an extensively leveraged weapon for software supply chain attacks.
  The detection of malicious packages is undoubtedly of high priority and many academic and commercial approaches have been developed.
  In the inevitable case of an attack, one needs resilience against malicious code.
  To this end, we present a runtime protection for Node.js that automatically limits a package's capabilities to an established minimum.
  The detection of required capabilities as well as their enforcement at runtime has been implemented and evaluated against known malicious attacks.
  Our approach was able to prevent 9/10 historic attacks with a median install-time overhead of less than 0.6 seconds and a median runtime overhead of less than 0.2 seconds.
\end{abstract}

%%
%% The code below is generated by the tool at http://dl.acm.org/ccs.cfm.
%% Please copy and paste the code instead of the example below.
%%


%%
%% Keywords. The author(s) should pick words that accurately describe
%% the work being presented. Separate the keywords with commas.

\received{20 February 2007}
\received[revised]{12 March 2009}
\received[accepted]{5 June 2009}

%%
%% This command processes the author and affiliation and title
%% information and builds the first part of the formatted document.
\maketitle

\section{Introduction}
\label{sec:introduction}
\input{sections/00-introduction.tex}

\section{Related Work}
\label{sec:related-work}
\input{sections/01-related-work.tex}

\section{Methodology}
\label{sec:methodology}
\input{sections/02-methodology.tex}

\section{Implementation}
\label{sec:implementation}
\input{sections/03-implementation.tex}

\section{Evaluation}
\label{sec:evaluation}
\input{sections/04-evaluation.tex}

\section{Discussion}
\label{sec:discussion}
\input{sections/05-discussion.tex}

\section{Conclusion \& Future Work}
\label{sec:conclusion-and-future-work}
\input{sections/06-conclusion-and-future-work.tex}

%%
%% The next two lines define the bibliography style to be used, and
%% the bibliography file.
% \clearpage
\balance
\bibliographystyle{ACM-Reference-Format}
\bibliography{bibliography.bib}

\end{document}

%% file: glossary.tex
\newacronym{AST}{AST}{Abstract Syntax Tree}
\newacronym{FOSS}{FOSS}{Free and Open Source Software}
\newacronym{ESM}{ESM}{ECMAScript Module}
\makeglossaries

%% file: sections/00-introduction.tex
Modern software lives and thrives from the opportunistic reuse of software components.
This is largely fueled by the sheer amount of \gls{FOSS}.
While the availability of ready-to-use building blocks certainly has its advantages, it also conveys a noticeable risk for security.
Effectively, each added software component increases the dependency on unverified code from untrusted developers.
Thus, it is no surprise that there is a visible trend for software supply chain attacks through maliciously manipulated software packages~\cite{ohm2020backstabber}.

Even when carefully choosing a software, the use of dependencies is nontransparent to the user.
A typical user does not notice if an attacker introduced malicious code somewhere down the supply chain.

In order to mitigate these supply chain attacks, we make use of two observations.
Firstly, the fact that a prominent attack vector is the introduction of malicious code at the patch level of a package that is automatically updated on the victim's system, and secondly, that it is observed that malicious additions noticeably alter the way a software component works~\cite{ohm2020towards}.
Making use of these facts, our approach is based on the continuity of benign software.
The overall goal is to prevent the execution of intentionally added malicious code which inherently requires other capabilities than usual.
To tackle the problem of using dependencies with unknown capabilities we propose an automated approach.

Our approach automatically infers required capabilities, like access to specific modules or functions, by statically analyzing the source code of a package as well as the source code of all of its dependencies.

In the first step, our approach infers a set of capabilities based on a trusted version of a package.
After updating that package to a newer version, it is run using our patched Node.js interpreter which enforces the established capabilities at runtime.
Newly added capabilities will not be accessible this way and hence corresponding code will not be executed successfully.

The presented approach is evaluated for caused overhead, exhaustiveness, possible attack surface reduction, and its performance on benign and malicious software.

The remainder of this paper is structured as follows.
First, we present and discuss related work in \Cref{sec:related-work}.
In \Cref{sec:methodology}, our methodology and use case are described and \Cref{sec:implementation} depicts the corresponding implementation.
This is followed by the presentation of results from our experiments in \Cref{sec:evaluation} and a discussion in \Cref{sec:discussion}.
Lastly, a conclusion is drawn and perspectives for future work are given in \Cref{sec:conclusion-and-future-work}.

%% file: sections/01-related-work.tex
% What broad fields do we touch?
% Securing the supply chain: some have tried to create general detection systems, some have tried
% update detection systems, some try to detect typosquatting.
% Resillience against malicious updates:
% - Installtime attacks
% - At runtime
% Capability based model
% - how to choose capabilities
% - how to enforce capabilities
With increasing amounts of attacks on the software supply chain in recent years, research in that field has been thriving.

Several works focused on providing security by identifying malicious software packages in package repositories.
This has been done both through code analysis~\cite{liang2021malicious, sejfia2022practical, ohm2022feasibility} and metadata analysis~\cite{zahan2021weak, gonzales2021anomalicious}.
In addition, Taylor et al.\ have built a tool that protects against typosquatting attacks~\cite{taylor2020defending}.
Further research has specialized this task to not just detect malicious packages in general, but instead aiming to detect malicious updates to previously benign packages, leveraging different kinds of anomaly detection~\cite{ohm2020towards, duan2019toward, garrett2019detecting}.
% In addition to straight up detecting malicious packages, there have also been approaches to detect differences between software releases, like binaries or packages within package repositories, and the corresponding open-sourced sourcecode~\cite{vu2021lastpymile, lamb2021reproducible}.
% This approach highlights the problem, that while open source software is in principle auditable, you often do not know whether the released version you're using actually corresponds to the published sourcecode.

However, even if the mentioned detectors were widely deployed, they would inevitably miss some malicious packages.
To account for this fact, there have also been approaches to allow the safe execution of malicious code.

Koishybayev and Kapravelos~\cite{koishybayev2020mininode} created a tool that reduces the attack surface of Node.js applications by removing unused code segments and blocking the ability to access built-in modules that are not statically referenced.
While this protects against attacks where dependencies abuse certain vulnerabilities in adjacent dependencies, it does not provide protection against malicious packages in general.

Vasilakis et al.~\cite{vasilakis2020breakapp} built an alternative module system for Node.js, which allows spawning instances of modules in configurable \textquote{compartments} providing different levels of isolation.
Only the compartment with the strongest isolation provides protection against malicious packages, but in turn requires extensive configuration for each module and introduces significant runtime overhead.

% Capability based resillience against install time attacks
In 2022, Wyss et al.~\cite{wyss2022wolf} have developed a way to protect users against install-time attacks.
Their system \emph{LATCH} is based around user-defined policies and aims to establish protection in two steps.
At first, it is determined whether a package should be installed.
To do so, a cloud service installs the suspected package in a sandbox and creates a manifest of so-called \textquote{intents}, based on the system call trace of the installation process.
If this manifest does not match the policy, it may optionally be blocked from installation on the user's machine.
Additionally, if the installation is performed, it is run within an AppArmor environment, which also enforces the adherence to the given policy.

While this approach seems to work well for the intended purpose, it has some problems.
At first, it puts the burden of creating a meaningfully secure policy on the user.
An attempt to mitigate this is to provide default policies.
However the evaluation results show that there are still cases where installations get blocked by the default policies, in which cases the user would have to adjust them.
Furthermore, this approach exclusively provides protection from install-time attacks, but does not provide any protection against runtime attacks, while current research suggests that more than 40~\% of software supply chain attacks are triggered at runtime~\cite{ohm2020backstabber}.

For protection against runtime attacks, Ferreira et al.~\cite{ferreira2021containing} have created a permission system with the intent of allowing individual packages to have individual permissions, which can be applied to packages of the Node.js ecosystem.
Package developers have to declare the permissions that their package needs, and the user has to accept the permissions given to a package.
At runtime, these permissions are enforced by restricting access to built-in JavaScript modules, as well as certain global objects.
Results are promising that capability-based approaches are able to reduce attack surface and capable of protecting against real-world attacks.
Finer grained capabilities would allow for an even larger proportion of the attack surface to be reduced.

The presented approaches rely on the user to create or verify the capability choice.
While they might be familiar with such a permission system from platforms like android, relying on the user to choose or verify permissions for a certain package bears risks.
Current research suggests that users do not pay attention to requests for certain permissions, and if they do, they do not fully understand the implications~\cite{felt2012android}.

For these reasons, we present a similar approach, leveraging more granular permissions and a system to automatically infer and enforce policies, without the need of a user to define the application's permissions.
In \Cref{sec:comparison} we will compare our approach and evaluation results to the work of Wyss et al.\ and Ferreira et al.\ where applicable.

%% file: sections/02-methodology.tex
Our main goal is to establish a system that automatically infers required capabilities of software and limits further access based on the principle of least privilege.
To this end, we first outline our general use case.
This is followed by the research statement providing the general procedure and research questions.
Last, we describe the experiments carried out in order to answer the research questions and thus substantiate our proposed solution.

\subsection{Use Case}~\label{sec:usecase}
We have chosen JavaScript and the Node.js runtime as our ecosystem, as it has been a common ecosystem choice for related work~\cite{sejfia2022practical, ohm2022feasibility, zahan2021weak, garrett2019detecting, wyss2022wolf, ferreira2021containing} and shows the highest number of known malicious packages~\cite{ohm2020backstabber}.
It thus meets our criteria of comparability to other research in the same area, as well as practical relevance.

We design our approach from the perspective of users who run some software on their computer \textemdash{} either for their own or hosted for someone else.
Consequentially, members of our target group are not necessarily developers but knowledgeable of software operating.
The key requirements for the user are running and updating the chosen software which shall provide uninterrupted productivity.

They actively selected the software in question and thus put a certain level of trust in the correctness (and implicitly benignity) of it.
That software again might have dependencies in form of other software components from which capabilities are leveraged to implement provided functionality.
Overall, the chain of software components is nontransparent to the user, and they might not notice changes in the dependencies and certainly will not track all changes made to the software itself.
Thus, while selecting the software, a one-sided trust relationship is established.

This relationship is regularly misused by attackers.
According to established taxonomies~\cite{ohm2020backstabber,ladisa2022taxonomy} they may manipulate an existing dependency in the software supply chain of the targeted software, add a new (malicious) dependency to the targeted software, or directly add new (malicious) functionality to the targeted software.
We assume that an attacker is able to identify the set of capabilities a software holds.
Without our approach they might choose any dependency of the software to be attacked.
When using our approach, they must select a software component suitable for the planned attack.
This heavily limits the attack surface.

We focus on the prevention of execution of malicious code despite how it was added to the targeted software.
However, we assume that the malicious code is contained in a future update of the software which the user is currently running.

\subsection{Research Statement}~\label{sec:research-statement}
    We develop an approach to automatically infer leveraged program functionalities, i.e., attributes and functions from global objects and built-in modules.

    The inferred capabilities need to be persisted to a policy file which will be used for the enforcement later on.
    To do so, we need to locate and understand the import functionality of the Node.js interpreter.
    Furthermore, that functionality needs to be enhanced in a way that allows it to respect our generated policy.
    The actual selections and details of the implementation will be presented in \Cref{sec:implementation}.

    We evaluate our approach to answer the following research questions:
    % \textbf{REPHRASE THE QUESTIONSSS}
    \begin{questions}
        \item How much overhead is added by the inference and enforcement of capabilities and policies respectively?~\label{itm:rq1}
        \item How exhaustive is the automated inference of capabilities?~\label{itm:rq2}
        \item How many capabilities are typically leveraged per package?~\label{itm:rq3}
        \item To what extent are benign package updates affected?~\label{itm:rq4}
        \item Which historic attacks would have been blocked?~\label{itm:rq5}
    \end{questions}
    \vspace{1em}

    Generally speaking, \ref{itm:rq1} evaluates the practical performance during operation and therefore is important to end users.

    Furthermore, \ref{itm:rq2} assesses how exhaustive, respectively, how complete, the inference of capabilities is.
    We measure how many capabilities are detected and how many are missed.

    The answer to \ref{itm:rq3} gauges to what extent our approach might reduce the attack surface of a software by limiting the access to (otherwise allowed) capabilities not listed in the policy.

    Assuming that most software updates are benign, we want to keep the amount of wrongly prevented code execution as small as possible.
    To this end, \ref{itm:rq4} estimates the approach's specificity by providing the expected amount of false positives (policy enforcement prevents benign software from execution) and true negatives (benign software is executed).

    The actual goal of our approach is to prevent the injection of malicious code in a software's updates.
    Thus, \ref{itm:rq5} determines the sensitivity by measuring true positives (policy enforcement prevents malicious software from execution) and false negatives (malicious software is executed).
    In order to answer these questions, we conduct several experiments.

\subsection{Experiments}~\label{sec:experiments}
    All experiments are run as \textquote{no-human-in-the-loop} style, i.e., all policies are generated and enforced fully automated.
    Nonetheless, our experiments provide an upper boundary for the approach's shortcomings and a lower boundary for its inference performance.

    To answer \ref{itm:rq1}, we perform five experiments on 200,000 randomly chosen packages.
    Regarding the first step, the capability inference, we measure the time it takes to infer all capabilities.
    Regarding the second step, the capability enforcement, we measure the time the source code replacement of global objects consumes, as well as the file sizes of the respective policies to estimate the impact on file load time.
    Additionally, we perform a theoretical analysis of the enforcement of module and global object restrictions.

    For \ref{itm:rq2} we use the 200,000 randomly chosen packages again, and compare the dependencies they use to the third party modules our approach is able to infer.

    As a reference of dependencies a package uses, we consult the list of runtime dependencies in the package's \texttt{package.json} manifest.
    Since the list of dependencies has to be manually maintained by the developer, it is prone to errors.
    For example, these lists can contain development dependencies incorrectly declared as runtime dependencies, or deprecated dependencies from previous versions that have not been removed from the dependency list.
    For this reason, we only consider those dependencies whose names also appear within at least one JavaScript file of the package.
    Additionally, we filter out all dependencies starting with \texttt{@types}, as those are just the type definitions needed to transpile code from TypeScript to JavaScript, and thus won't appear in JavaScript files.

    In order to quantify how much our approach reduces the allowed capabilities of a program, we enumerate the set of generally available capabilities.
    We then calculate the fraction of actually required capabilities of 200,000 random packages.
    This will answer \ref{itm:rq3} by providing the numbers of available and actually required capabilities.

    The change of required capabilities caused by historic updates is recorded for \ref{itm:rq4}.
    We conducted the experiment on the 1,000 most depended upon software packages from npm because these are most likely to be benign.
    This allows us to provide an upper bound of false positives and true negatives of our approach.

    For a set of known malicious packages taken from the Backstabber's Knife Collection~\cite{ohm2020backstabber}, we generate and enforce the policy on the preceding (last benign) version of the affected package and trigger the malicious behavior of the malicious sample in a sandboxed environment.
    This allows us to determine whether we are able to prevent this attacks to answer \ref{itm:rq5}.
    Furthermore, a qualitative analysis of preventions is carried out to gain more insight.

    All experiments are conducted twice.
    Once for the coarse granularity at module level and once for the finer granularity at member level (c.f. \Cref{sec:implementation}).
    The results of the experiments are presented in \Cref{sec:evaluation}.

    The source code of our approach as well as the software packages' names used for the experiments are available on GitHub\footnote{\url{https://github.com/cybertier/npm-dependency-guardian}}.

%% file: sections/03-implementation.tex
After explaining the general concept of our approach, this chapter presents our reference implementation.
We detail the implementation of the policy generation and show how the generated policy is enforced at runtime.

\subsection{Policy Generation}
The first step of the implementation is the generation of the policy.
Our policy is a mapping of a package name to its capabilities.
Therefore, we first have to make a concrete choice what exactly our capabilities are.

As described by Ferreira et al., Node.js has very limited functionality without importing any of its built-in modules~\cite{ferreira2021containing}.
Additionally, the modules are inherently grouping abilities that belong together.
For example, the \texttt{fs} module allows access to different kinds of file system operations, like reading or writing files.
Therefore, we are selecting the built-in modules as one part of the capabilities.

Furthermore, JavaScript and the Node.js interpreter expose a set of global objects which also allow access to certain abilities.
For example, the \texttt{Buffer} global object allows manipulation of byte buffers, and allows performing various en- and decoding operations.
As we are aiming to provide a more fine granular approach than Ferreira et al., we are also considering global objects to be part of the capabilities.
Thus, the set of capabilities we are choosing for our implementation is the union of the set of built-in modules and the set of global objects.

We also evaluate the feasibility of different capability granularities.
For that purpose, we are using our previously selected capability set as the coarse-grained capability set.
Since these coarse-grained capabilities are usually groups of different operations, as previously shown at the example of the \texttt{fs} module, the fine-grained capability set will consist of the individual operations within these groups, meaning the members of the module objects.

To identify a package's capabilities, we generate an \gls{AST} for each file within the package with one of the JavaScript file extensions \texttt{.mjs}, \texttt{.cjs} or \texttt{.js}, using \emph{acorn}~\cite{acornjs}.
Each file that acorn fails to parse is considered invalid JavaScript and is thus ignored.

Given the \gls{AST} we extract the used modules and global objects in the coarse-grained setting, and additionally their members in the fine-grained setting.
Imported modules are identified as the arguments to calls to either the \texttt{require} or \texttt{import} function or the \texttt{import} statement.
Global objects are identified as identifiers satisfying the following three properties:
\begin{enumerate}
    \item Its name is the name of a global object
    \item It references an object in the current scope
    \item Its name was not overwritten by a local variable
\end{enumerate}

In the fine-grained setting, we additionally extract the accessed members of the imported modules and accessed global objects.
Member accesses can happen through either array- or object-patterns on the module- or global-objects, or via member expressions.
All of these explicitly list the accessed member in the respective \gls{AST} node.

% An example for the module inference is visualized in~\Cref{fig:ast}.

% \begin{figure}
%   \centering
%   \includegraphics[width=\columnwidth]{figs/ast.pdf}
%   \caption{Source code (top right) as well as a simplified corresponding \gls{AST}. Highlighted subgraphs represent the levels of granularity for the inference of accessed built-in modules \textemdash{} coarse-grained (black/solid) and fine-grained (blue/dashed).}~\label{fig:ast}
% \end{figure}

After extracting the capabilities for every JavaScript file of a package, the package's capabilities are set to the union of all those files' capabilities.
The policy contains all the packages in the dependency graph with their corresponding capabilities.
In the coarse-grained case, these are just the imported modules and used global objects.
The fine-grained policy contains all the capabilities of the coarse-grained policy, and additionally contains all the accessed members of the imported modules and used global objects.
An example snippet of a policy is shown in~\Cref{fig:example-policy}.
Since Node.js is natively able to parse JSON, the policy is stored in the JSON format.

\begin{figure}
  \begin{lstlisting}[language=json,firstnumber=1]
{
  "memberAccessTracing": true,
  "policyCoarse": {
    "ast-package": {
      "modules": ["fs"],
      "globals": ["require"]
    }
  },
  "policyFine": {
    "ast-package": {
      "modules": ["fs.readFile"],
      "globals": []
    }
  }
}
  \end{lstlisting}
  \caption{Example policy. If \texttt{memberAccessTracing} is \texttt{false}, then \texttt{policyFine} may be empty.}~\label{fig:example-policy}
\end{figure}

\subsection{Policy Enforcement}
The generated policy is enforced through a patch in the Node.js runtime.
The enforcement is split into two parts.

The first part is the enforcement of the module restrictions.
This is done by patching the \texttt{makeRequire} function in the file \texttt{lib/internal/modules/cjs/helpers.js}, which is responsible for providing the \texttt{require} function to every loaded module.\footnote{Policy enforcement for ECMAScript modules is currently not supported as further discussed in \Cref{subsubsec:runtimeenforcement}.}

In the coarse-grained model, we replace every module that is not contained in the requiring packages' allowlist with a dummy object containing the same members as the required module, but where the corresponding value is a function that returns itself when it's called.
This way, a package that is not allowed to access a certain module can still require it and call its member functions without crashing, but with no risk of performing malicious actions through that module.

In the fine-grained model we return a copy of the module, where only those members that are not present in the policy are replaced with our dummy function, while the members that are contained in the policy still point to the original module functions.
Additionally, if the module itself is callable, it will execute the original function if the module itself is contained in the policy.
If the module is not contained in the policy, calling it will execute our dummy function.

The second part is the enforcement of global object restrictions.
To do this, we inject a new local object into each module, which holds a reference to all global objects that are present in the allowlist.
Similar to the module enforcement, references to objects that are not part of the policy will be replaced with dummy objects.

Additionally, when the file is loaded, we alter the source code, replacing each reference to a global object with a reference to the corresponding member of our new injected local object.
This way, we can ensure that only those global objects present in the allowlist can actually be accessed.

Alterations of this process for the fine-grained model happen analogously to the module restriction enforcement.

%% file: sections/04-evaluation.tex
In this section we present and briefly discuss the results of our experiments as listed in \Cref{sec:experiments}.

\subsection{Overhead (\ref{itm:rq1})}
In order to understand the caused overhead, there are two spots to consider.
First, the capability inference \textit{before} the runtime and second the policy enforcement \textit{during} the runtime.
The capability inference time is measured on 200,000 randomly chosen packages from npm as discussed in \Cref{sec:experiments}.
We separately measured our approach with and without member access tracing enabled in order to see its impact on the overhead.

\begin{figure}[thb]
  \centering
  \includegraphics[width=\columnwidth]{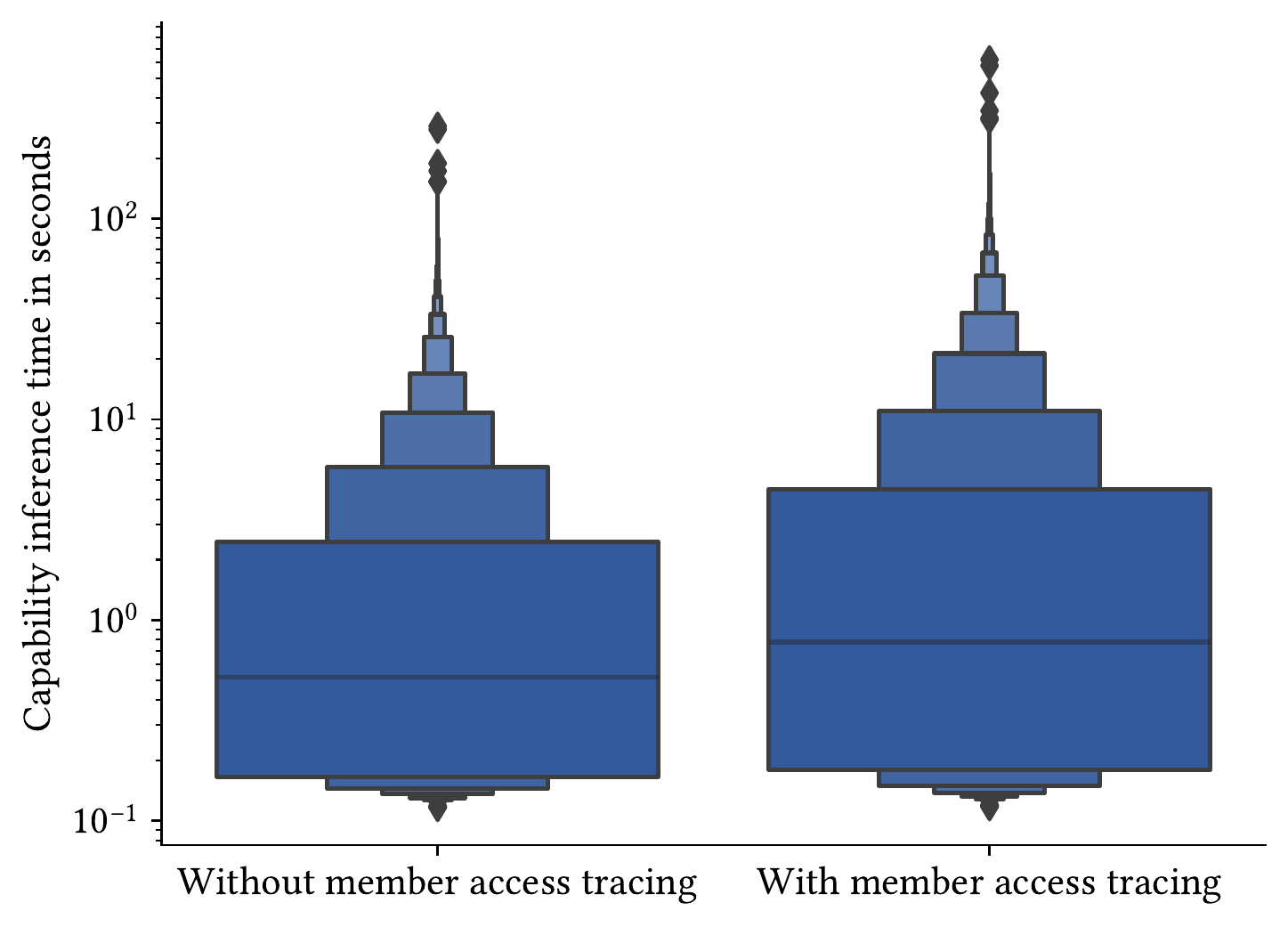}
  \caption{Letter-value plot~\cite{hofmann2011letter} of capability inference times in seconds for 192,546 randomly chosen packages from npm. Please note the logarithmic scale.}~\label{fig:overhead}
\end{figure}
In \Cref{fig:overhead} one can see two letter-value plots~\cite{hofmann2011letter} (logarithmic scale) of the capability inference times in seconds with and without member access tracing enabled.

On average, the policy generation took 2.73 seconds without member access tracing and 5.21 seconds with member access tracing.
It yielded a standard deviation of 6.08 and 12.32 respectively.
The median is at 0.52 and 0.60 and thus much lower than the average.
Correspondingly, we found out that 75~\% of the policies can be generated within 2.45 or 4.50 seconds and 90~\% can be generated within 7.29 and 14.10 seconds.
Even the slowest 1~\% can be generated within 31.07 and 62.88 seconds.
The absolute slowest generation took 288.54 seconds without member access tracing and 618.49 seconds with member access tracing and hence constitutes as a heavy outlier.

We were unable to evaluate about 8,200 packages due to installation errors, either because npm has reported an exception when trying to install the package, or was not able to complete the installation within a five-minute time window.

Furthermore, it should be noted that the policy generation is performed only once for each version of a software.
Thus, the overhead is added at the very first installation and at every update of a software.

% Scanned 191836 packages. Member access tracing: False
% Low  1% 0.12503894567489623
% Low 10% 0.1412191390991211
% Low 25% 0.1660900115966797
% Median: 0.5208276510238647
% High 25% 2.45383620262146
% High 10% 7.2968361377716064
% High  1% 31.068965089321136
% Mean: 2.7334272544895897
% Min: ('@malware-test-selfs-jujus-fores-picul/test-mlw3-selfs-jujus-fores-picul', {'dir': '@malware-test-selfs-jujus-fores-picul-test-mlw3-selfs-jujus-fores-picul-1.0.1-security.0-7b2a17e5-1e6a-4162-863a-a904a9e4984a', 'scan_time': 0.1163630485534668})
% Max: ('sol-bin', {'dir': 'sol-bin-3.2.1-bb650cf5-5693-4d0a-939c-780ddbaefb3e', 'scan_time': 288.5443150997162})
% Stddev: 6.076311820209345

% Scanned 191836 packages. Member access tracing: True
% Low  1% 0.1264062762260437
% Low 10% 0.14478623867034912
% Low 25% 0.1796908974647522
% Median: 0.7781524658203125
% High 25% 4.498674929141998
% High 10% 14.101611733436584
% High  1% 62.883082604408266
% Mean: 5.213665073274977
% Min: ('dsr-rollback-package-kerfs-sadhu-joeys-unify', {'dir': 'dsr-rollback-package-kerfs-sadhu-joeys-unify-1.0.0-c5049962-5cbb-43e0-a4b3-34b933bce716', 'scan_time': 0.11743426322937012})
% Max: ('sol-bin', {'dir': 'sol-bin-3.2.1-ccabc1db-a65a-479b-b1b7-43967fdbe81d', 'scan_time': 618.4871528148651})
% Stddev: 12.323186632946943

Assuming that a well written program imports all dependencies at the beginning of the file, our policy enforcement solely causes overhead at the program start.
This is because our patched \texttt{require} function is called once for each import and returns the modified module for further use.
In case of additional module imports during the runtime, the policy enforcement would be triggered again.

Either way, policy enforcement of module restrictions, i.e., the pruning of the imported modules according to the precomputed policy, only requires two lookups in an in-memory hash map and \textemdash{} if appropriate \textemdash{} the replacement of module functions whenever a new import happens.

Restricting the access to global objects is a bit more evolved and requires some preparation as described in \Cref{sec:implementation}.
It requires the generation of an \gls{AST}, the identification of used global objects, and the dynamic replacement of global object usage within the source code.
We measured the added time to a program's startup by performing global object replacement for a whole software, i.e., all JavaScript files within a package and all its dependencies, on the 200,000 randomly chosen packages.
On median this procedure adds 198~ms whereas 90~\% of the packages are prepared within 2.67~s.
However, there are some heavy outliers that require 59.15~s of preparation time.

Per file our approach requires 165~\textmu{}s on median, while 90~\% of the files are processed within 1.12~ms.
Global replacement adds a measurable amount of overhead at the startup, but it is most often reasonable.

% Package stats
% Scanned 192002 datapoints
% Low  1% 0.00
% Low 10% 6,154,496.00
% Low 25% 23,898,442.50
% Median: 198,712,228.00
% High 25% 996,353,763.25
% High 10% 2,671,173,304.90
% High  1% 11,353,941,824.38
% Mean: 984,568,026.80
% Min: 0.00
% Max: 59,154,616,083.00
% Stddev: 2,105,520,178.85
%
% File stats
% Scanned 315093156 datapoints
% Low  1% 33,219.00
% Low 10% 53,937.00
% Low 25% 85,203.00
% Median: 165,216.00
% High 25% 429,428.00
% High 10% 1,115,917.00
% High  1% 7,305,038.45
% Mean: 599,946.48
% Min: 13,297.00
% Max: 11,162,625,113.00
% Stddev: 3,243,238.99

Loading a policy from disk to memory greatly depends on the file's size.
Thus, we calculated file sizes for the policies of the 200,000 randomly sampled packages.
Without member access tracing a policy file is 1.2~kB in size while 90~\% of the files are below 27.69~kB.
Even the absolute maximum of 399.57~kB should not have a noticeable impact on the program's startup time.

Policies that include member access tracing are larger in file size as they have to convey more information, and are supersets of the policies without member access tracing.
The median is at 2.98~kB and 90~\% of the files are below 65.10~kB.
The absolute maximum increased to 985.08~kB which still poses no issue.

% MEMBER TRACED:
% Statistics for 191654 packages
% Low  1% 176.0
% Low 10% 214.0
% Low 25% 327.0
% Median: 2980.0
% High 25% 18847.75
% High 10% 65104.1
% High  1% 342858.55
% Mean: 24252.252945412045
% Min: 164
% Max: 985084
% Stddev: 58229.91393188075
% Max policy for pkg @wildcards/harberger-ui 985084

% NON MEMBER TRACED:
% Statistics for 191654 packages
% Low  1% 88.0
% Low 10% 105.0
% Low 25% 156.0
% Median: 1202.0
% High 25% 8080.0
% High 10% 27694.7
% High  1% 147369.64
% Mean: 10320.107167082346
% Min: 82
% Max: 399570
% Stddev: 24913.229690206354
% Max policy for pkg @wildcards/harberger-ui 399570
Thus, its total overhead is negligible.
Overall, we conclude that our approach adds a reasonable amount of overhead that is outweighed by its security gains.

\begin{rqanswer}
  \textbf{Response to \ref{itm:rq1}:} On median our approach requires 0.52 (0.60) seconds to generate a policy. The actual enforcement of the policy adds a negligible overhead of 198~ms to the program's start. Overall, our approach introduces only a small footprint.
\end{rqanswer}

\subsection{Exhaustiveness (\ref{itm:rq2})}\label{subsec:exhaustiveness}
To analyze the exhaustiveness, i.e., does our approach comprehensively infer the set of required capabilities, we performed the experiment as described in \Cref{sec:experiments}.
Recall that we compare the set of the detected external modules to the dependencies used by the package.

% Missing 41951 of 541538 dependencies.
% 17206 of 178787 have missing dependencies
% 69589 packages have no dependencies
% npm reg error 3311
% stack exceeded 12

Again, we conducted the experiment on the 200,000 randomly chosen packages from npm.
However, 3,311 packages had corrupt meta information and 12 crashed during \gls{AST} generation.
An additional 17,706 packages had invalid JavaScript files, for example incorrectly named \texttt{.jsx} files.
From the remaining packages we collected 541,538 dependencies in total.
Our tool correctly detected 499,587 (92.25~\%) dependencies.
Of the 109,198 packages with at least one dependency, we correctly inferred all dependencies for 91,992 (84.24~\%) packages.

As mentioned in \Cref{sec:experiments}, the validity of these results depends on the correct declaration of dependencies.
Even our naive verification process of looking for the dependency name within the package's JavaScript files is not guaranteed to only result in dependencies that are actually used.
Therefore, we manually inspected 100 randomly selected packages where at least one dependency was counted as undetected by our automated approach, to estimate the quality of this dataset.
We found that 80 of those packages did not actually use the declared runtime dependency, and the text matches were usually found as variable names or within comments.
In the remaining 20 cases we were not able to confirm that the respective module is imported, but from the given occurrence we could also not confidently deny that the module may be imported through some indirections.

This highlights that even with our additional verification measures, there are a lot of dependencies that are not actually used.
However, it is unlikely that there are dependencies that are used but not declared in the \texttt{package.json} file, as this would in most cases break the package for all users.
Therefore, we consider our findings as lower bounds for the amount of correctly found modules.

\begin{rqanswer}
  \textbf{Response to \ref{itm:rq2}:}
  Our approach correctly detects 92~\% of used imports, and for 84~\% of packages all imports are correctly detected.
\end{rqanswer}

\subsection{Reducibility of Capabilities (\ref{itm:rq3})}
The reduction of permitted code also reduces the attack surface of the software~\cite{ponta2021used}.
Thus, we conducted an experiment to measure how many of the available capabilities are not used by a package.
To this end, we counted how many capabilities are available to the set of 200,000 randomly samples packages and how many of those capabilities are not included in the package's policy, i.e., not required to run the software.

On median 78.82~\% of the available capabilities are not used when counting without member access tracing.
If member access tracing is enabled, the number of unused capabilities went up to 96.56~\%.
Furthermore, 99~\% of the packages require less than 80.59~\% of the available capabilities without member access tracing and 53.15~\% with member access tracing.

In order to put these numbers into perspective, let's take a look at the Node.js built-in module \texttt{fs}.
In total, it offers 101 members, e.g., functions to write, read, or append a file.
If one wants to write a file, exactly one member which is called \texttt{fs.writeFile} is required.
Consequentially, a large amount of members is unused and hence can be removed without breaking the program.

% \begin{figure}[thb]
%   \centering
%   \includegraphics[width=\columnwidth]{figs/attack-surface-reduction.pdf}
%   \caption{Fractions of unused capabilities per package when compared to the set of available capabilities.}~\label{fig:attack-surface-reduction}
% \end{figure}

% Scanned 192114 packages. Member access tracing: False
% Low  1% 0.0
% Low 10% 0.5882352941176471
% Low 25% 2.941176470588235
% Median: 21.176470588235293
% High 25% 43.529411764705884
% High 10% 61.17647058823529
% High  1% 80.58823529411765
% Mean: 25.874554262818215
% Min: 0.0
% Max: 95.29411764705881
% 
% Stddev: 23.59920626792208
% Scanned 192114 packages. Member access tracing: True
% Low  1% 0.0
% Low 10% 0.0
% Low 25% 0.2152080344332855
% Median: 3.443328550932568
% High 25% 11.406025824964132
% High 10% 22.309899569583933
% High  1% 53.15638450502152
% Mean: 8.046397875699371
% Min: 0.0
% Max: 97.99139167862266
% Stddev: 11.322683451566965

In conclusion, our approach drastically shrinks the available capabilities.
If an attacker wants to perform a certain action that requires a certain set of capabilities, they would either need to add these capabilities to the program's policy or infiltrate a project that already includes all the required capabilities.

\begin{rqanswer}
  \textbf{Response to \ref{itm:rq3}:} On median our approach may reduce the capabilities of a software by 78.82~\% (without member access tracing) and 96.56~\% (with member access tracing).
\end{rqanswer}

\subsection{Specificity (\ref{itm:rq4})}\label{subsec:specificity}
Being able to automatically infer, describe, and enforce capabilities for a software, we will now investigate its performance on benign software updates.
Software is updated rather frequently and thus our approach must be able to let updates happen without breaking the application we want to use.
Consequentially, we investigated how the set of capabilities per package changes across historic updates.

As we do not have a large dataset of known benign packages across all their versions, we conducted this experiment for the 1,000 most depended upon packages from npm according to \Cref{sec:experiments}, assuming that all currently published versions of these packages are benign.
Furthermore, we distinguish between major, minor, and patch updates according to Semantic Versioning~\cite{preston2013semantic} as well as if the updates required new modules, new global objects, or both in order to get more granular insights.

\begin{figure*}[thb]
  \includegraphics[width=0.9\textwidth]{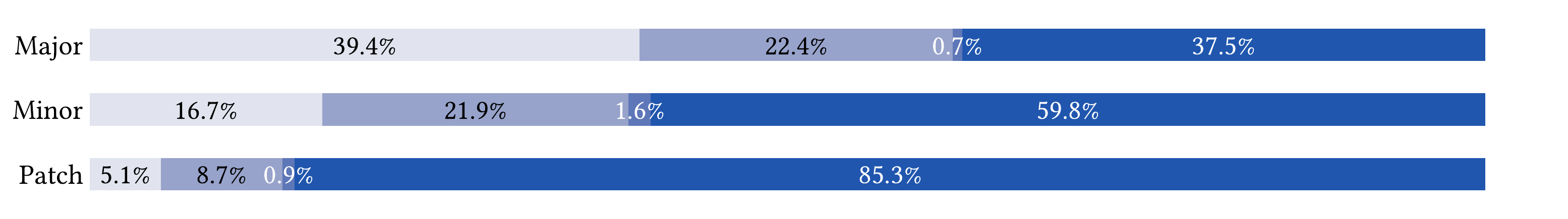}
  \caption{Changes of capabilities required \emph{without} member access tracing, sorted by version level.}~\label{fig:fp-wo-mt}
  \begin{center}
    \vspace{2em}
    \includegraphics[width=0.7\textwidth]{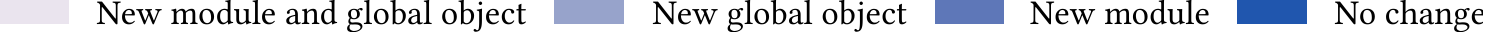}
  \end{center}
\end{figure*}
\begin{figure*}
  \includegraphics[width=0.9\textwidth]{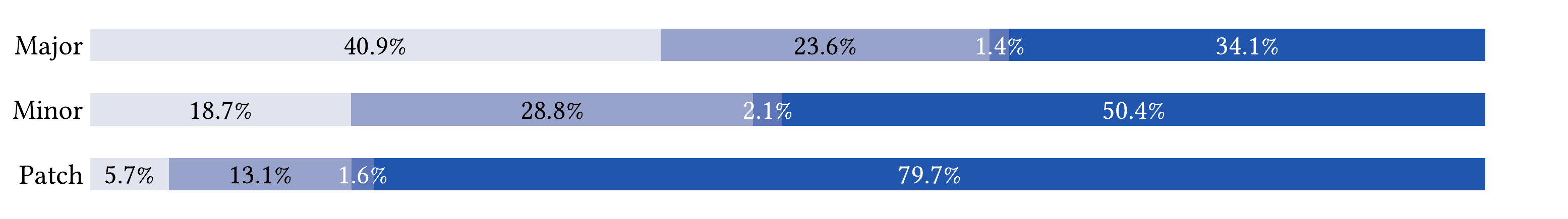}
  \caption{Changes of capabilities required \emph{with} member access tracing, sorted by version level.}~\label{fig:fp-w-mt}
\end{figure*}

As depicted in \Cref{fig:fp-wo-mt} and \Cref{fig:fp-w-mt}, 37.5~\% (34.1~\%)\footnote{Numbers in brackets represent results with enabled member access tracing.} of major updates did not require new capabilities.
However, if new capabilities were required 39.4~\% (40.9~\%) were caused by previously unused modules \emph{and} global objects.
Therefore, 22.4~\% (23.6~\%) of the updates solely used new global objects and merely 0.7~\% (1.4~\%) solely used new modules.

For minor updates, 59.8~\% (50.4~\%) of the updates did not require a change of granted capabilities.
Breaking down the newly required capabilities, one can see that 21.9~\% (28.8~\%) were caused by previously unused global objects.
This fraction roughly stayed the same as compared to major updates.
However, 1.6~\% (2.1~\%) were caused by the use of new modules.

The vast majority of updates \textemdash{} 85.3~\% (79.7~\%) \textemdash{} did not introduce a change of capabilities in patch updates.
This high number is to be expected as patch level updates should solely concern bug fixes.
Thus, they typically do not require new functionality and hence no new capabilities.
Still, 14.7~\% (20,4~\%) of the updates required new modules or global objects.

The amount of newly required capabilities drastically reduces from major over minor to patch updates.
Overall, the measured changes are in line with the procedure as proposed by Semantic Versioning~\cite{preston2013semantic}.

The conducted experiment yields an upper bound for the specificity, i.e., false positives and true negatives, of our approach.
Each newly required capability would be prohibited by our approach.
Arguably, major updates should not be performed fully automatically as it is expected to be not backwards compatible and hence might break things.

Most often, patch updates do not require new capabilities.
However, as observed by Ohm et al.~\cite{ohm2020towards}, malicious code is frequently introduced on patch level \textemdash{} a commonly allowed level for automatic updates \textemdash{} and alters the program's functionality.
Our approach would impede such an attack.
The evaluation against known historic attacks is discussed in the following section.

Moreover, it is not clear from the experiment if a newly required capability, which hence can not be used, would actually break the application.
It might be the case that this capability is required by a component down in the dependency tree and never required for the functionality by the actual application we are running.
\begin{rqanswer}
  \textbf{Response to \ref{itm:rq4}:} Applying our approach to patch updates is still likely to result in too many false positives. This motivates further improvements.
\end{rqanswer}

\subsection{Sensitivity (\ref{itm:rq5})}
Arguably, the most important question is~\ref{itm:rq5}, whether the presented approach protects against attacks, i.e., malicious updates.
According to our usecase described in \Cref{sec:usecase}, we are considering previously benign software packages that turn malicious.
This may be achieved by adding malicious code to the package itself, adding malicious code to one of its dependencies, or by adding a new dependency containing malicious code~\cite{ladisa2022taxonomy}.

As there exists no method of formally proving that we will be able to prevent all future attacks, we evaluate whether we would have prevented past attacks on benign packages.
To do so, we need samples of malicious packages that were infected in one of the presented manners.
The Backstabber's Knife Collection~\cite{ohm2020backstabber} is a curated collection of such samples.
It also has a well maintained package index, which allows querying individual packages by several metadata attributes.
Therefore, we use the Backstabber's Knife Collection\footnote{We are using the commit \texttt{22bd768}, which was the most up-to-date commit at the time of our evaluation.} to gather samples for our evaluation.

Packages appropriate for our attack model have to meet the following criteria:
\begin{enumerate}
  \item Published in the npm registry
  \item Infected an existing package
  \item Malicious action at runtime
  \item Applicable to Node.js
\end{enumerate}
Using the package index metadata mentioned above, we can create a query that preselects only packages that meet the first three criteria.
However, not all of them are applicable to Node.js, as some of them leverage browser exclusive JavaScript APIs to perform their malicious actions.
This characteristic is not reflected in the dataset's metadata and hence we removed such packages by hand.
These remaining packages as well as the malicious version and the last benign version for reference are listed in \Cref{tab:b2m-versions}.
\begin{table}[thb]
  \caption{List of package versions used for our experiment, if the attack was averted, and if the program flow crashed because of our interference.}~\label{tab:b2m-versions}
  \begin{tabularx}{\columnwidth}{lrcc}
    \toprule
    \textbf{Package}       & \textbf{Update}            & \textbf{Averted} & \textbf{Crashed} \\
    \midrule
    conventional-changelog & 1.1.24 $\rightarrow$ 1.2.0 & \ding{51}        & \ding{55}        \\
    eslint-config-eslint   & 5.0.1 $\rightarrow$ 5.0.2  & \ding{51}        & \ding{55}        \\
    eslint-scope           & 3.7.1 $\rightarrow$ 3.7.2  & \ding{51}        & \ding{55}        \\
    event-stream           & 3.3.5 $\rightarrow$ 3.3.6  & \ding{51}        & \ding{55}        \\
    kraken-api             & 0.1.7 $\rightarrow$ 0.1.8  & \ding{51}        & \ding{51}        \\
    leetlog                & 0.1.1 $\rightarrow$ 0.1.2  & \ding{51}        & \ding{51}        \\
    load-from-cwd-or-npm   & 3.0.1 $\rightarrow$ 3.0.2  & \textbf{?}       & \textbf{?}       \\
    mariadb                & 2.5.6 $\rightarrow$ 2.13.0 & \ding{51}        & \ding{51}        \\
    opencv.js              & 1.0.0 $\rightarrow$ 1.0.1  & \ding{51}        & \ding{51}        \\
    rate-map               & 1.0.2 $\rightarrow$ 1.0.3  & \ding{51}        & \ding{51}        \\
    \bottomrule
  \end{tabularx}
\end{table}

To evaluate whether our approach would have prevented an attack, we create a policy for the respective preceding version of the package and afterwards trigger the malicious behavior of the malicious version in a sandboxed environment.
We will breifly discuss each sample in regards to if and why it averted the attack.

\subsubsection*{conventional-changelog}
This package decodes a command with the \texttt{from} member of the \texttt{Buffer} global object to a command that is then passed to the \texttt{spawn} member of the \texttt{child\_process} module, which downloads and executes a cryptocurrency miner.
Neither of these modules and members had been used in the benign version.
Thus, we prevent this attack and do so without interupting the programs runtime, i.e., no crashing.

\subsubsection*{eslint-config-eslint \& eslint-scope}
The malicious function downloads a payload using the member \texttt{get} of the module \texttt{http} which is then executed using \texttt{eval}.
Again, neither of the modules nor members had been used previously, and the attack would have been prevented without interruption of the runtime.

\subsubsection*{event-stream}
For this well-known sample, the actual malicious code is present in the newly added dependency \emph{flatmap-stream}.
This piece of code first decrypts some AES encrypted data with the \texttt{createDecipher} member of the \texttt{crypto} module, which results in a string containing JavaScript code.
This code is then evaluated using the \texttt{constructor} method of the \texttt{module} global object.
As the \emph{flatmap-stream} dependency was newly added to the dependencies with the malicious update of \emph{event-stream}, there is no policy entry for it yet, thus it does not have access to any modules or global objects.
Therefore, the attack would have been prevented without crashing the program.

\subsubsection*{kraken-api}
In this case, a reverse-shell would open up using the modules \texttt{net.socket} and \texttt{child\_process.spawn}.
None of these were used in the benign version and hence the attack was averted by our runtime protection.
However, the program crashed because a module called \texttt{daemon}, used to run the reverse-shell in the background, could not be imported succesfully due to missing capabilities.

\subsubsection*{leetlog}
The malicious update uses the \texttt{readDir} and \texttt{appendFile} members of the \texttt{fs} module to add an ssh-key to the \texttt{authorized\_keys} file.
The module \texttt{fs} had not been used in the benign version.
Thus, this attack would have also been prevented but the program crashed.

\subsubsection*{load-from-cwd-or-npm}
This attack aimed at one of its dependents, the \emph{purescript-installer} package, by returning a \texttt{PassThrough} object from the \texttt{stream} module, resulting in an endless loop in the purescript-installer.
\texttt{stream} had not been used before, and thus it would yield our dummy object instead of the \texttt{PassThrough} object.
While it is likely that this would not have resulted in an endless loop, it would probably still have led to a crash of the purescript-installer.
Since we were not able to trigger the malicious behavior, we do not consider this attack as being prevented.

\subsubsection*{maria-db \& opencv.js}
Both attacks tried to send the environment variables to a server controlled by the attacker by using \texttt{querystring.stringify} and \texttt{http.request}.
For both packages these modules and respective members had never been used before.
In the case of \texttt{opencv.js}, the member \texttt{process.env} was also unused.
Thus, in both cases the attack was averted by our runtime protection but both times causing the program to crash.

\subsubsection*{rate-map}
This attack uses the \texttt{fs} module and its members \texttt{readFileSync}, \texttt{writeFileSync} and \texttt{existSync}, as well as the global object \texttt{Buffer.from} to rewrite the source code of an adjacent module, and remove its own malicious part of the code.
The policy of the previous benign version contains no modules.
It does contain some global objects, but not \texttt{Buffer} or any of its members.
Therefore, this attack would have been prevented, but our runtime protection caused the program's crash.

In conclusion, the proposed approach is able to avert 9/10 of the investigated attacks.
Our naive implementation of the mock object which is returned if a prohibited capability is requested turned out to be insufficient in most of the cases.
We also observed that using the fine granular policy does not provide any benefit in stopping the given attacks, as none of them used the required modules or global objects in their benign versions.

\begin{rqanswer}
  \textbf{Response to \ref{itm:rq5}:} Our approach would have stopped 9 out of 10 known attacks. Member access tracing does not yield better results.
\end{rqanswer}

\subsection{Comparison to Related Work}~\label{sec:comparison}
Our approach stands in direct comparison with Wyss et al.~\cite{wyss2022wolf} and Ferreira et al.~\cite{ferreira2021containing}.
While Wyss et al.\ focus on install-time attacks, we focus on runtime protection.
In contrast to Ferreira et al.\, who leverage a manually defined and coarse capability set, we take a more granular approach based on automatically inferred policies.
Nonetheless, some metrics can be compared.

% policy generation
The policies in the approach of Ferreira et al.\ are declared by the developers in the coarse categories \textit{network access}, \textit{file system access}, \textit{process creation}, and \textit{all}.
Wyss et al.\ and us use automatically inferred policies based on built-in modules and globals.
Their policy inference takes less than one minute for 90~\% of the packages.
Our approach requires 7.25 seconds (without member access tracing) and 14.10 seconds (with member access tracing) for 90~\% of the packages.

% overhead
Ferreira et al.\ measure an overhead of less than 1~\% during runtime of 20 common CLI tools.
Wyss et al.\ found out that their approach can enforce their policy in less than a second for 99~\% of the packages.
Our approach adds 198~ms to the program's start and a negligible amount of time during runtime.

% benign dataset
In order to get an impression on the false positves the approaches are tested with benign packages.
The approach of Wyss et al.\ blocks 1.5~\% of all packages using a generic and predefined policy.
Ferraira et al.\ did not analyze the performance on benign packages.
Our approach is evaluated on the 1000 most depended upon packages and revealed that only 14.7~\% of patch update would require an update of the policy.

% malicious dataset
All approaches are tested against known attacks.
Ferreira et al.\ replayed three attacks and were able to avert all of them.
Wyss et al.\ analyze 102 samples taken from the Backstabber's Knife Collection, the same source we leveraged.
They crafted the policies to match common install-time attacks observed in these samples and hence are able to detect all of them.
We executed 9 known malicious packages on policies generated on the previous benign version and were able to avert all the attacks.

% misc
Ferreira et al.\ estimate to protect 14--33~\% of the available packages as their do not require any permissions.
Our experiments show that on median 78.82~\% (96.56~\% with member access tracing) of the available capabilities are indeed unused.

In comparison, our approach seems to be fast and precise.
Due to the more granular concept of our policies, we tend to be too sensitive and may also deny benign updates.

%% file: sections/05-discussion.tex
We will first discuss limitations to our general concept and then follow up with a discussion about our concrete implementation, as well as our evaluations.

\subsection{Conceptual}\label{subsec:discussionconceptual}
Our approach is based on the assumption that required capabilities of a software do not change over its versions.
The evaluation of \ref{itm:rq4} in \Cref{subsec:specificity} has shown that this assumption does not hold in its entirety.
However, we have seen that narrowing down the assumption to be specific to only patch updates, the number of new capabilities falls drastically.
While this may not be the ideal choice yet, more alternatives for a more specific assumption should be explored, for example altering the concrete choice for the considered capabilities.

The choice of all modules and global objects as equal capabilities makes sense for a generic use case.
However, when thinking about malicious packages, some modules may prove more useful than others.
For example, the \texttt{fs} module is useful for common use cases like data exfiltration or overwriting sensitive files, while the \texttt{v8} module for interacting with certain APIs of the underlying V8 runtime may be less useful for an attacker.
Furthermore, the \texttt{process} module provides the ability to create arbitrary processes on the system.
As these processes are usually not subject to our restricted Node.js interpreter, they essentially have all capabilities, meaning that any module with the \texttt{process} capability can do everything.
Using a more elaborate approach for the capability choice, for example based on an analysis of their use in malware or their perceived usefulness for it, may thus lead to increased performance.

Regarding the detection of capabilities, our choice of static analysis also brings expected limitations, as it is not able to cover dynamic aspects.
Related literature like Vasilakis et al.~\cite{vasilakis2021preventing} suggest that modules will not dynamically need what they do not need statically.

\subsection{Proof of Concept Implementation}
Besides conceptual points, there are also points to discuss in our proof of concept implementation.
To this end, we are differentiating between the implementation of the policy generation, and the policy enforcement.

\subsubsection{Difference to Node.js Built-In Policies}
Node.js comes with its own experimental implementation of policies~\cite{nodejspolicies} allowing a per-module restriction and redirection of module imports.
While these policies allow to implement parts of our approach, they are missing some necessary features.

Most notably, they only support the restriction of module imports, but cannot be used to restrict global objects at all.
Furthermore, replacing a module with a mock module requires a separate file containing the specific mock module, as well as a separate entry in the policy for every distinct pair of importing module and imported module, while our approach only lists the allowed capabilities, and automatically mocks all capabilities that are not explicitly listed.
This is especially problematic, as all pairs that are not explicitly listed will not be replaced with a mock module, but will instead lead to a crash of the program or no action at all \textemdash{} depending on the default action setting in the policy \textemdash{} which we explicitly attempt to prevent.

For these reasons, we have chosen to design our own policy format, as well as our own policy enforcement.

\subsubsection{Policy Generation}
We decided to solely analyze files which have a JavaScript file extension (\texttt{.mjs}, \texttt{.cjs} or \texttt{.js}) for the generation of the \gls{AST}.
However, Node.js allows the execution and import of files with arbitrary file extensions.

Aside from that, we do not include \emph{bundle dependencies} in our scan, even though they are supported by npm.
The reason is that they are not included in the \texttt{package-lock.json} file, thus we do not have a mapping of file paths to package names, which we need to add them to our policy.
Since only two of the top 1000 most depended upon packages and 0.38~\% of the whole npm registry are using bundle dependencies, we do not see this as an urgent issue.
However, bundle dependencies could be added in the future, at the cost of a more complex dependency discovery algorithm.

\subsubsection{Policy Enforcement}\label{subsubsec:runtimeenforcement}
There are some ways an attacker could circumvent our proposed protection mechanism.

Firstly, as our policy enforcement of module restrictions only alters the \texttt{require} function, it does not support \glspl{ESM}, and thus any \gls{ESM} has access to any module.
While only about 3.23\% of all packages on npm are \glspl{ESM}, they account for 13.3\% of the 1000 most depended upon packages.
Therefore, it is desirable to expand the implementation and support \glspl{ESM} in the future.

Furthermore, there are more ways to import code in JavaScript than the standard \texttt{import} and \texttt{require} functions.
For our proof of concept we focused on these standard procedures.
Handling the alternative types of importing code can be added in the future.

At last, all our protection mechanisms only work for native JavaScript modules.
However, Node.js also allows the use of C++ extensions~\cite{nodejscppextensions} which do not need to access any modules or global objects, thus circumventing our mechanism in the same way an entirely new process does.

\subsection{Evaluation}\label{subsec:discussionevaluation}
As all of our evaluations were automatically installing packages from the npm registry, we were subject to regular erroneous behavior shown by the npm CLI\@.
We have ignored all packages that npm did not manage to install, or whose installation did not finish within five minutes, as there exist several packages that will set npm in a state of infinite dependency resolution.
However, one might argue that such an installation time limit may bias the distribution of successful installation towards smaller packages, as larger packages with more dependencies naturally take longer to install.
We argue that any working package should be able to install within five minutes.

Regarding \ref{itm:rq3}, our reference for all the available properties is only a lower bound for the available capabilities when using member access tracing, as JavaScript does not provide a way to reliably enumerate all the members of an object~\cite{mdnproperties}.
Our count of the available capabilities might thus be lower than the actual amount of available capabilities.

%% file: sections/06-conclusion-and-future-work.tex
Software supply chain attacks that utilize maliciously manipulated software modules are on the rise.
A good portion of these inject malicious code directly into a software or more obfuscated in its dependencies.
This code is eventually executed during runtime at the end user's machine.

In this paper we present and evaluate a system that automatically infers and enforces capabilities for software based on the principle of least privilege.
By doing so, we prevent the execution of untrusted code.

For our approach, a capability is understood as access to a certain module of a software or a member of it, e.g., a class and its provided attributes.
Through static analysis, we determine the minimal set of modules that are required for a software to run.
This information is persisted to a policy file in form of an allowlist.
That policy is used by our approach to enforce the access to capabilities.
To this end, imported modules are trimmed to its allowed members according to the policy just in time.
Our approach supports two modes: fine-grained (with member access tracing) and coarse-grained (without member access tracing).

We conduct several experiments on software packages available from the npm package repository for Node.js/JavaScript to validate our approach.

The use of our system has a small footprint, i.e., fast generation of policies (0.52~s/0.60~s) and small file size (1.2~kB/2.98~kB) for persistence.
Furthermore, it adds only a slight overhead of 198~ms to the program's startup time.

Our overall naive approach leverages Abstract Syntax Trees and detects 92~\% of a program's capabilities.
However, it still causes too many false positives and requires further engineering.
Still, our research indicates the general viability of the approach.
Moreover, we would have been able to prevent at least 9 out of 10 historic attacks.
Once a policy is generated it can be used for any future version of a software.
Benign updates tend to be unaffected by this rigorous allowlist, but malicious updates often required an extended set of capabilities and hence failed to execute.

Furthermore, we found out that 78.82~\%  of available capabilities in the coarse-grained mode and by 96.56~\% in the fine-grained mode are indeed unused.
Thus, our approach potentially reduces the attack surface by this share.
This means that an attacker would need to infiltrate a package that already has the capabilities required for the malicious actions.
Accordingly, attacks on the software supply chain that need to alter the way a software component operates will be prevented.

While we successfully demonstrated our concept and a corresponding proof of concept, there is still room for improvements.
A lot of shortcomings stem from the use of Node.js/JavaScript for our experiments.
The functionality of it is so flexible that there always seems to be another way to circumvent our implementation.
Nonetheless, our approach may be adopted to other languages easily.

For future work we would like to further improve our approach and transfer it to other languages such as Python.
Furthermore, the weighting of capabilities, for instance \texttt{eval} and \texttt{child\_process} may more easily be misused, or some fuzzy logic for \textquote{very unusual imports} may yield better results.

%% file: paper.bbl
%%% -*-BibTeX-*-
%%% Do NOT edit. File created by BibTeX with style
%%% ACM-Reference-Format-Journals [18-Jan-2012].

\begin{thebibliography}{24}

%%% ====================================================================
%%% NOTE TO THE USER: you can override these defaults by providing
%%% customized versions of any of these macros before the \bibliography
%%% command.  Each of them MUST provide its own final punctuation,
%%% except for \shownote{}, \showDOI{}, and \showURL{}.  The latter two
%%% do not use final punctuation, in order to avoid confusing it with
%%% the Web address.
%%%
%%% To suppress output of a particular field, define its macro to expand
%%% to an empty string, or better, \unskip, like this:
%%%
%%% \newcommand{\showDOI}[1]{\unskip}   % LaTeX syntax
%%%
%%% \def \showDOI #1{\unskip}           % plain TeX syntax
%%%
%%% ====================================================================

\ifx \showCODEN    \undefined \def \showCODEN     #1{\unskip}     \fi
\ifx \showDOI      \undefined \def \showDOI       #1{#1}\fi
\ifx \showISBNx    \undefined \def \showISBNx     #1{\unskip}     \fi
\ifx \showISBNxiii \undefined \def \showISBNxiii  #1{\unskip}     \fi
\ifx \showISSN     \undefined \def \showISSN      #1{\unskip}     \fi
\ifx \showLCCN     \undefined \def \showLCCN      #1{\unskip}     \fi
\ifx \shownote     \undefined \def \shownote      #1{#1}          \fi
\ifx \showarticletitle \undefined \def \showarticletitle #1{#1}   \fi
\ifx \showURL      \undefined \def \showURL       {\relax}        \fi
% The following commands are used for tagged output and should be
% invisible to TeX
\providecommand\bibfield[2]{#2}
\providecommand\bibinfo[2]{#2}
\providecommand\natexlab[1]{#1}
\providecommand\showeprint[2][]{arXiv:#2}

\bibitem[Duan(2019)]%
        {duan2019toward}
\bibfield{author}{\bibinfo{person}{Ruian Duan}.} \bibinfo{year}{2019}\natexlab{}.
\newblock \emph{\bibinfo{title}{Toward solving the security risks of open-source software use}}.
\newblock \bibinfo{thesistype}{Ph.\,D. Dissertation}. \bibinfo{school}{Georgia Institute of Technology}.
\newblock


\bibitem[Felt et~al\mbox{.}(2012)]%
        {felt2012android}
\bibfield{author}{\bibinfo{person}{Adrienne~Porter Felt}, \bibinfo{person}{Elizabeth Ha}, \bibinfo{person}{Serge Egelman}, \bibinfo{person}{Ariel Haney}, \bibinfo{person}{Erika Chin}, {and} \bibinfo{person}{David Wagner}.} \bibinfo{year}{2012}\natexlab{}.
\newblock \showarticletitle{Android Permissions: User Attention, Comprehension, and Behavior}. In \bibinfo{booktitle}{\emph{Proceedings of the Eighth Symposium on Usable Privacy and Security}} (Washington, D.C.) \emph{(\bibinfo{series}{SOUPS '12})}. \bibinfo{publisher}{Association for Computing Machinery}, \bibinfo{address}{New York, NY, USA}, Article \bibinfo{articleno}{3}, \bibinfo{numpages}{14}~pages.
\newblock
\showISBNx{9781450315326}


\bibitem[Ferreira et~al\mbox{.}(2021)]%
        {ferreira2021containing}
\bibfield{author}{\bibinfo{person}{Gabriel Ferreira}, \bibinfo{person}{Limin Jia}, \bibinfo{person}{Joshua Sunshine}, {and} \bibinfo{person}{Christian K{\"a}stner}.} \bibinfo{year}{2021}\natexlab{}.
\newblock \showarticletitle{Containing malicious package updates in npm with a lightweight permission system}. In \bibinfo{booktitle}{\emph{2021 IEEE/ACM 43rd International Conference on Software Engineering (ICSE)}}. IEEE, \bibinfo{pages}{1334--1346}.
\newblock


\bibitem[Garrett et~al\mbox{.}(2019)]%
        {garrett2019detecting}
\bibfield{author}{\bibinfo{person}{Kalil~Anderson Garrett}, \bibinfo{person}{Gabriel Ferreira}, \bibinfo{person}{Limin Jia}, \bibinfo{person}{Joshua Sunshine}, {and} \bibinfo{person}{Christian K{\"{a}}stner}.} \bibinfo{year}{2019}\natexlab{}.
\newblock \showarticletitle{Detecting suspicious package updates}. In \bibinfo{booktitle}{\emph{Proceedings of the 41st International Conference on Software Engineering: New Ideas and Emerging Results, {ICSE} {(NIER)} 2019, Montreal, QC, Canada, May 29-31, 2019}}, \bibfield{editor}{\bibinfo{person}{Anita Sarma} {and} \bibinfo{person}{Leonardo Murta}} (Eds.). \bibinfo{publisher}{{IEEE} / {ACM}}, \bibinfo{pages}{13--16}.
\newblock


\bibitem[Gonzalez et~al\mbox{.}(2021)]%
        {gonzales2021anomalicious}
\bibfield{author}{\bibinfo{person}{Danielle Gonzalez}, \bibinfo{person}{Thomas Zimmermann}, \bibinfo{person}{Patrice Godefroid}, {and} \bibinfo{person}{Max Schaefer}.} \bibinfo{year}{2021}\natexlab{}.
\newblock \showarticletitle{Anomalicious: Automated Detection of Anomalous and Potentially Malicious Commits on GitHub}. In \bibinfo{booktitle}{\emph{43rd {IEEE/ACM} International Conference on Software Engineering: Software Engineering in Practice, {ICSE} {(SEIP)} 2021, Madrid, Spain, May 25-28, 2021}}. \bibinfo{publisher}{{IEEE}}, \bibinfo{pages}{258--267}.
\newblock


\bibitem[Heine~né Lang et~al\mbox{.}({[n.\,d.]})]%
        {acornjs}
\bibfield{author}{\bibinfo{person}{Adrian Heine~né Lang}, \bibinfo{person}{Marijn Haverbeke}, {and} \bibinfo{person}{Ingvar Stepanyan}.} \bibinfo{year}{[n.\,d.]}\natexlab{}.
\newblock \bibinfo{title}{acorn}.
\newblock
\newblock
\urldef\tempurl%
\url{https://github.com/acornjs/acorn}
\showURL{%
\tempurl}
\newblock
\shownote{[Last accessed on 2022-10-18]}.


\bibitem[Hofmann et~al\mbox{.}(2011)]%
        {hofmann2011letter}
\bibfield{author}{\bibinfo{person}{Heike Hofmann}, \bibinfo{person}{Karen Kafadar}, {and} \bibinfo{person}{Hadley Wickham}.} \bibinfo{year}{2011}\natexlab{}.
\newblock \bibinfo{booktitle}{\emph{Letter-value plots: Boxplots for large data}}.
\newblock \bibinfo{type}{{T}echnical {R}eport}. \bibinfo{institution}{had.co.nz}.
\newblock


\bibitem[Koishybayev and Kapravelos(2020)]%
        {koishybayev2020mininode}
\bibfield{author}{\bibinfo{person}{Igibek Koishybayev} {and} \bibinfo{person}{Alexandros Kapravelos}.} \bibinfo{year}{2020}\natexlab{}.
\newblock \showarticletitle{Mininode: Reducing the Attack Surface of Node.js Applications}. In \bibinfo{booktitle}{\emph{23rd International Symposium on Research in Attacks, Intrusions and Defenses, {RAID} 2020, San Sebastian, Spain, October 14-15, 2020}}, \bibfield{editor}{\bibinfo{person}{Manuel Egele} {and} \bibinfo{person}{Leyla Bilge}} (Eds.). \bibinfo{publisher}{{USENIX} Association}, \bibinfo{pages}{121--134}.
\newblock


\bibitem[Ladisa et~al\mbox{.}(2023)]%
        {ladisa2022taxonomy}
\bibfield{author}{\bibinfo{person}{P. Ladisa}, \bibinfo{person}{H. Plate}, \bibinfo{person}{M. Martinez}, {and} \bibinfo{person}{O. Barais}.} \bibinfo{year}{2023}\natexlab{}.
\newblock \showarticletitle{SoK: Taxonomy of Attacks on Open-Source Software Supply Chains}. In \bibinfo{booktitle}{\emph{2023 2023 IEEE Symposium on Security and Privacy (SP) (SP)}}. \bibinfo{publisher}{IEEE Computer Society}, \bibinfo{address}{Los Alamitos, CA, USA}, \bibinfo{pages}{167--184}.
\newblock


\bibitem[Liang et~al\mbox{.}(2021)]%
        {liang2021malicious}
\bibfield{author}{\bibinfo{person}{Genpei Liang}, \bibinfo{person}{Xiangyu Zhou}, \bibinfo{person}{Qingyu Wang}, \bibinfo{person}{Yutong Du}, {and} \bibinfo{person}{Cheng Huang}.} \bibinfo{year}{2021}\natexlab{}.
\newblock \showarticletitle{Malicious Packages Lurking in User-Friendly Python Package Index}. In \bibinfo{booktitle}{\emph{20th {IEEE} International Conference on Trust, Security and Privacy in Computing and Communications, TrustCom 2021, Shenyang, China, October 20-22, 2021}}. \bibinfo{publisher}{{IEEE}}, \bibinfo{pages}{606--613}.
\newblock


\bibitem[{Mozilla Corporation}({[n.\,d.]})]%
        {mdnproperties}
\bibfield{author}{\bibinfo{person}{{Mozilla Corporation}}.} \bibinfo{year}{[n.\,d.]}\natexlab{}.
\newblock \bibinfo{title}{Enumerability and ownership of properties - JavaScript | MDN}.
\newblock
\newblock
\urldef\tempurl%
\url{https://developer.mozilla.org/en-US/docs/Web/JavaScript/Enumerability_and_ownership_of_properties}
\showURL{%
\tempurl}
\newblock
\shownote{[Last accessed on 2022-10-20]}.


\bibitem[Ohm et~al\mbox{.}(2022)]%
        {ohm2022feasibility}
\bibfield{author}{\bibinfo{person}{Marc Ohm}, \bibinfo{person}{Felix Boes}, \bibinfo{person}{Christian Bungartz}, {and} \bibinfo{person}{Michael Meier}.} \bibinfo{year}{2022}\natexlab{}.
\newblock \showarticletitle{On the Feasibility of Supervised Machine Learning for the Detection of Malicious Software Packages}. In \bibinfo{booktitle}{\emph{Proceedings of the 17th International Conference on Availability, Reliability and Security}}. \bibinfo{pages}{1--10}.
\newblock


\bibitem[Ohm et~al\mbox{.}(2020a)]%
        {ohm2020backstabber}
\bibfield{author}{\bibinfo{person}{Marc Ohm}, \bibinfo{person}{Henrik Plate}, \bibinfo{person}{Arnold Sykosch}, {and} \bibinfo{person}{Michael Meier}.} \bibinfo{year}{2020}\natexlab{a}.
\newblock \showarticletitle{Backstabber’s knife collection: A review of open source software supply chain attacks}. In \bibinfo{booktitle}{\emph{International Conference on Detection of Intrusions and Malware, and Vulnerability Assessment}}. Springer, \bibinfo{pages}{23--43}.
\newblock


\bibitem[Ohm et~al\mbox{.}(2020b)]%
        {ohm2020towards}
\bibfield{author}{\bibinfo{person}{Marc Ohm}, \bibinfo{person}{Arnold Sykosch}, {and} \bibinfo{person}{Michael Meier}.} \bibinfo{year}{2020}\natexlab{b}.
\newblock \showarticletitle{Towards detection of software supply chain attacks by forensic artifacts}. In \bibinfo{booktitle}{\emph{{ARES} 2020: The 15th International Conference on Availability, Reliability and Security, Virtual Event, Ireland, August 25-28, 2020}}, \bibfield{editor}{\bibinfo{person}{Melanie Volkamer} {and} \bibinfo{person}{Christian Wressnegger}} (Eds.). \bibinfo{publisher}{{ACM}}, \bibinfo{pages}{65:1--65:6}.
\newblock


\bibitem[{OpenJS Foundation}(2022)]%
        {nodejscppextensions}
\bibfield{author}{\bibinfo{person}{{OpenJS Foundation}}.} \bibinfo{year}{2022}\natexlab{}.
\newblock \bibinfo{title}{C++ addons | Node.js v18.11.0 Documentation}.
\newblock
\newblock
\urldef\tempurl%
\url{https://nodejs.org/api/globals.html}
\showURL{%
\tempurl}
\newblock
\shownote{[Last accessed on 2022-10-18]}.


\bibitem[{OpenJS Foundation}(2023)]%
        {nodejspolicies}
\bibfield{author}{\bibinfo{person}{{OpenJS Foundation}}.} \bibinfo{year}{2023}\natexlab{}.
\newblock \bibinfo{title}{Permissions | Node.js v18.15.0 Documentation}.
\newblock
\newblock
\urldef\tempurl%
\url{https://nodejs.org/api/permissions.html}
\showURL{%
\tempurl}
\newblock
\shownote{[Last accessed on 2023-03-08]}.


\bibitem[Ponta et~al\mbox{.}(2021)]%
        {ponta2021used}
\bibfield{author}{\bibinfo{person}{Serena~Elisa Ponta}, \bibinfo{person}{Wolfram Fischer}, \bibinfo{person}{Henrik Plate}, {and} \bibinfo{person}{Antonino Sabetta}.} \bibinfo{year}{2021}\natexlab{}.
\newblock \showarticletitle{The Used, the Bloated, and the Vulnerable: Reducing the Attack Surface of an Industrial Application}. In \bibinfo{booktitle}{\emph{{IEEE} International Conference on Software Maintenance and Evolution, {ICSME} 2021, Luxembourg, September 27 - October 1, 2021}}. \bibinfo{publisher}{{IEEE}}, \bibinfo{pages}{555--558}.
\newblock


\bibitem[Preston-Werner(2013)]%
        {preston2013semantic}
\bibfield{author}{\bibinfo{person}{Tom Preston-Werner}.} \bibinfo{year}{2013}\natexlab{}.
\newblock \bibinfo{title}{Semantic Versioning 2.0.0}.
\newblock
\newblock
\urldef\tempurl%
\url{https://semver.org/lang/de/}
\showURL{%
\tempurl}
\newblock
\shownote{[Last accessed on 2022-10-17]}.


\bibitem[Sejfia and Sch{\"{a}}fer(2022)]%
        {sejfia2022practical}
\bibfield{author}{\bibinfo{person}{Adriana Sejfia} {and} \bibinfo{person}{Max Sch{\"{a}}fer}.} \bibinfo{year}{2022}\natexlab{}.
\newblock \showarticletitle{Practical Automated Detection of Malicious npm Packages}. In \bibinfo{booktitle}{\emph{44th {IEEE/ACM} 44th International Conference on Software Engineering, {ICSE} 2022, Pittsburgh, PA, USA, May 25-27, 2022}}. \bibinfo{publisher}{{ACM}}, \bibinfo{pages}{1681--1692}.
\newblock


\bibitem[Taylor et~al\mbox{.}(2020)]%
        {taylor2020defending}
\bibfield{author}{\bibinfo{person}{Matthew Taylor}, \bibinfo{person}{Ruturaj~K. Vaidya}, \bibinfo{person}{Drew Davidson}, \bibinfo{person}{Lorenzo~De Carli}, {and} \bibinfo{person}{Vaibhav Rastogi}.} \bibinfo{year}{2020}\natexlab{}.
\newblock \showarticletitle{Defending Against Package Typosquatting}. In \bibinfo{booktitle}{\emph{NSS}}.
\newblock


\bibitem[Vasilakis et~al\mbox{.}(2018)]%
        {vasilakis2020breakapp}
\bibfield{author}{\bibinfo{person}{Nikos Vasilakis}, \bibinfo{person}{Ben Karel}, \bibinfo{person}{Nick Roessler}, \bibinfo{person}{Nathan Dautenhahn}, \bibinfo{person}{Andr{\'{e}} DeHon}, {and} \bibinfo{person}{Jonathan~M. Smith}.} \bibinfo{year}{2018}\natexlab{}.
\newblock \showarticletitle{BreakApp: Automated, Flexible Application Compartmentalization}. In \bibinfo{booktitle}{\emph{25th Annual Network and Distributed System Security Symposium, {NDSS} 2018, San Diego, California, USA, February 18-21, 2018}}. \bibinfo{publisher}{The Internet Society}.
\newblock


\bibitem[Vasilakis et~al\mbox{.}(2021)]%
        {vasilakis2021preventing}
\bibfield{author}{\bibinfo{person}{Nikos Vasilakis}, \bibinfo{person}{Cristian{-}Alexandru Staicu}, \bibinfo{person}{Grigoris Ntousakis}, \bibinfo{person}{Konstantinos Kallas}, \bibinfo{person}{Ben Karel}, \bibinfo{person}{Andr{\'{e}} DeHon}, {and} \bibinfo{person}{Michael Pradel}.} \bibinfo{year}{2021}\natexlab{}.
\newblock \showarticletitle{Preventing Dynamic Library Compromise on Node.js via RWX-Based Privilege Reduction}. In \bibinfo{booktitle}{\emph{{CCS} '21: 2021 {ACM} {SIGSAC} Conference on Computer and Communications Security, Virtual Event, Republic of Korea, November 15 - 19, 2021}}, \bibfield{editor}{\bibinfo{person}{Yongdae Kim}, \bibinfo{person}{Jong Kim}, \bibinfo{person}{Giovanni Vigna}, {and} \bibinfo{person}{Elaine Shi}} (Eds.). \bibinfo{publisher}{{ACM}}, \bibinfo{pages}{1821--1838}.
\newblock


\bibitem[Wyss et~al\mbox{.}(2022)]%
        {wyss2022wolf}
\bibfield{author}{\bibinfo{person}{Elizabeth Wyss}, \bibinfo{person}{Alexander Wittman}, \bibinfo{person}{Drew Davidson}, {and} \bibinfo{person}{Lorenzo De~Carli}.} \bibinfo{year}{2022}\natexlab{}.
\newblock \showarticletitle{Wolf at the Door: Preventing Install-Time Attacks in npm with Latch}. In \bibinfo{booktitle}{\emph{Proceedings of the 2022 ACM on Asia Conference on Computer and Communications Security}}. \bibinfo{pages}{1139--1153}.
\newblock


\bibitem[Zahan et~al\mbox{.}(2021)]%
        {zahan2021weak}
\bibfield{author}{\bibinfo{person}{Nusrat Zahan}, \bibinfo{person}{Laurie~Ann Williams}, \bibinfo{person}{Thomas Zimmermann}, \bibinfo{person}{Patrice Godefroid}, \bibinfo{person}{Brendan Murphy}, {and} \bibinfo{person}{Chandra~Shekhar Maddila}.} \bibinfo{year}{2021}\natexlab{}.
\newblock \showarticletitle{What are Weak Links in the npm Supply Chain?}
\newblock \bibinfo{journal}{\emph{ArXiv}}  \bibinfo{volume}{abs/2112.10165} (\bibinfo{year}{2021}).
\newblock


\end{thebibliography}
